
\def\d#1/d#2{ {\partial #1\over\partial #2} }



\def\pdr{\partial}

\def\al{\alpha}

\def\de{\delta}

\def\half{{1\over 2}}
\def\tr{\hbox{tr}}

\def\un#1{\underline{#1}}
\def\Ad{\hbox{Ad}}


\newcount\eqnumber
\def\beq{ \global\advance\eqnumber by 1 $$ }
\def\eeq{ \eqno(\the\eqnumber)$$ }
\def\n{\global\advance \eqnumber by 1\eqno(\the\eqnumber)}
\def\puteqno{
\global\advance \eqnumber by 1 (\the\eqnumber)}
\def\beqs{$$\eqalign}
\def\eeqs{$$}


\def\ifundefined#1{\expandafter\ifx\csname
#1\endcsname\relax}
 \newcount\sectnumber \sectnumber=0
\def\sect#1{ \advance \sectnumber by 1 {\it \the \sectnumber. #1} }

\newcount\refno \refno=0  
\def\[#1]{
\ifundefined{#1}
\advance\refno by 1
\expandafter\edef\csname #1\endcsname{\the\refno}\fi[\csname
#1\endcsname]}
\def\refis#1{\noindent\csname #1\endcsname. }

\def\label#1{
\ifundefined{#1}
\expandafter\edef\csname #1\endcsname{\the\eqnumber}
\else\message{label #1 already in use}
\fi{}}
\def\(#1){(\csname #1\endcsname)}
\def\eqn#1{(\csname #1\endcsname)}

\baselineskip=15pt
\parskip=10pt
\magnification=1200
\def\BEGINIGNORE#1ENDIGNORE{}

\baselineskip=20pt

\def\t#1{\tilde{#1}}
\def\cA{ {\cal A} }
\def\cF{\cal F}

\def\ym{ Yang--Mills }
\def\tW{\tilde{W}}


\def\D{ {d\over dx} }
\def\cA{ {\cal A} }
\def\cG{ {\cal  G} }
\def\t#1{ \tilde{#1} }
\def\ad{ \hbox{ad}\; }
\def\lie{ {\cal L} }

				\hfill November 1993

                                \hfill UR-1327

                                \hfill ER-40685-777\break

\vskip.1in
\centerline{\bf Yang--Mills Theory on a Cylinder Coupled to
Point Particles
 }
\vskip.1in
\centerline{\rm   K. S. Gupta, R.J. Henderson,  S. G. Rajeev and
O.T. Turgut}
\vskip.1in
\centerline{\it Department of Physics and Astronomy}
\centerline{\it University of Rochester}
\centerline{\it Rochester, N.Y. 14627}
\centerline{\it e-mail: gupta,henderson,rajeev and
 turgut@urhep.pas.rochester.edu}
\vskip.4in

\baselineskip=24 true pt

\centerline{\bf Abstract}

We study a model of quantum Yang--Mills theory with a finite number of gauge
invariant degrees of freedom. The gauge field has only a finite number of
degrees of freedom since we assume that space--time is a two dimensional
cylinder. We couple the gauge field to matter, modeled by either one or two
nonrelativistic point particles. These problems can be solved {\it without
any gauge fixing}, by generalizing the canonical quantization methods of
Ref.\[rajeev] to the case including matter. For this, we make  use of the
geometry of the space of connections, which has the structure of a Principal
Fiber Bundle with an infinite dimensional fiber.  We are able to reduce both
problems to finite dimensional, exactly solvable, quantum mechanics problems.
In the case of one particle, we find that the ground state energy will diverge
in the limit of infinite radius of space, consistent with confinement. In the
case of two particles, this does not happen if they can form a color singlet
bound state (`meson').

\vfill\eject

\sect{ Introduction}

Understanding nonabelian gauge theories at the quantum level
is a problem of
fundamental significance in  particle physics as well as
mathematical physics.
 In particle physics this is related to the as yet unresolved
problem of
deducing a theory of hadrons from Quantum Chromodynamics.
There has been much
progress in the two dimensional Quantum
Chromodynamics\[hadron] which does
not have any true gauge degrees of freedom. Studying two
dimensional gauge
theories on spaces with nontrivial topology
\[migdal],\[rajeev],\[hosotani],\[witten],\[blauthompson]
is interesting since there are a finite number of gauge
degrees of freedom.
This allows us to study the dynamics of quantum gauge theories
nonperturbatively in a context where the problem is exactly
solvable. The
insights learned from this solution should be useful in more
realistic
situations.

In a more mathematical direction, recall that classical gauge
theories\[isham],\[percacci] have a natural formulation in
terms of Principal
Fiber Bundles. The dynamical variable of gauge theory is a
connection in a
Principal Fiber bundle. The fundamental symmetry
transformations of the
theory, gauge transformations, form the ( fiber preserving) group of
 automorphisms
of the Principal
Fiber Bundle.
 Matter fields coupled to the gauge field are described in
terms of sections of
associated bundles. In the quantum theory, we are interested
not in a
particular connection , but rather in the structure of the
space of all
connections. The group of gauge transformations act on it; two
connections that
differ only by a gauge transformation  are to be considered
equivalent. It is
remarkable fact that this situation also has a natural
description in terms of
an infinite dimensional  Principal Fiber
Bundle\[narasimhan],\[singer],\[babelon],\[mickbook].   The space of
connections is the total space and the base space
is the quotient space of gauge equivalent connections. Thus
the proper
geometric setting for quantum gauge theories is this Principal
Fiber Bundle.
For example, the wavefunctions of a gauge theory  are sections
of an
associated vector bundle. Even when  the original ( classical)
Principal bundle
is trivial this quantum bundle is nontrivial. Choosing a
section of this
bundle is equivalent to choosing a gauge. The fact that there
is no global
section for the quantum bundle implies that there is no choice
of gauge valid
for all configurations: this situation is often referred to as
the `Gribov
ambiguity' in the physics literature \[gribov].
Moreover, on this infinite dimensional fiber bundle, there is
a natural
connection. The Kinetic Energy of the gauge field is the
covariant Laplacian
with respect to this connection. This leads to an additional
coupling between
the matter and gauge sectors which is not physically obvious.
( This  is in
addition to the more physically obvious Coulomb interaction.)

In most cases of direct physical interest, this covariant
Laplacian cannot yet
be given a precise meaning: the divergences of quantum field
theory have to be
overcome. We will study a very simple class of systems which
have only a finite
number of ( gauge invariant) degrees of freedom. Therefore it
will be possible
in the end to define   this operator,  although the steps
leading upto it
involve formal infinite dimensional manipulations. The gauge
field has no
propagating ( `wave--like') degrees of freedom if the space--
time is 1+1
dimensional. Still if there are non--contractible loops, there
will be a finite
number of  gauge invariant degrees of freedom associated to the
parallel
transport ( `Wilson loop') around these loops. The simplest
case \[rajeev]  is
the one where space--time is a cylinder, and gauge theory
without matter can
then be reduced to a simple problem in quantum mechanics with
a finite number
of degrees of freedom. It is possible to couple this theory to
matter fields,
\[hosotani],\[mickelsson], \[langmann]
which will have an infinite number of degrees of freedom
associated to matter.
This makes the problem too complicated to be solved exactly. A
simpler model,
more in the spirit of \[rajeev] is to couple gauge theory to
sources which are
nonrelativistic point particles. ( One can view this as the
limiting case
where the matter consists of heavy particles.) This theory
has only a finite number of
degrees of freedom and can be solved by reducing it to a
quantum mechanical
problem again. This is the approach we  will follow in this
paper.  We believe
that the  detailed study of such simple systems will help the
develop the
intuition  and technology to be applied to more complicated
problems. In fact
one lesson we have learned from this exercise is that it is
best to follow a
manifestly gauge invariant approach, taking into account that
the bundle of
connections is nontrivial. We  hope to  apply the lessons
learned from
these simple models to more complicated problems such as 2+1
dimensional  gauge
theories.

There has been a revival of interest in exactly solvable gauge
theories in 1+1
dimensions\[ho],\[hetrick],\[shabanov],\[ferrari],\[saradzhev]
.  Some of this work relates this problem to topological field
theories \[thompsonreview],\[broda]. There is also an
interesting new approach to the quotient
space of connections,\[eliza]  which we hope will be
generalized to the case with matter fields. A mathematically
rigorous path integral approach to two dimensional gauge
theories has  also been developed \[fine]. The exact
solvability of the path integral can be understood in terms of
localization \[wittenlocal].

Now we will summarize the contents of this paper. In  section
2 we review some
basic ideas on Principal Fiber Bundles. In the finite
dimensional case this is
standard material. We do not as yet have enough examples to
develop a rigorous
theory for infinite dimensional bundles. Instead, we will use
the ideas
familiar from the finite dimensional case in a formal way in
the infinite
dimensional context.   In fact, one of the motivations for
our work is to understand what the correct definitions should
be, in a more rigorous approach. In section  3, the geometry
of the bundle
of connections is developed in more detail, in the special
case  of gauge
theory on a cylinder. The  simplifications of this  special
case are used
throughout to obtain as explicit a description as possible.
For example, the
base space ( the quotient space of connections modulo gauge
transformations) is
finite dimensional. The projection map of the bundle can be
constructed as the
solution of  an ordinary differential equation.
 We   describe a flat Riemann metric on the space of
connections, invariant
under gauge transformations. This induces a curved metric on
the base space.
The vertical vector fields on the bundle describe
infinitesimal gauge
transformations. We can now  define a horizontal vector to be
orthogonal to all
the vertical vectors. This establishes a connection on the
quantum  bundle.  In section
4  we discuss an associated vector bundle of this principal
bundle.  A
characterization of a section of the associated vector bundle
as
quasi--periodic function on the real line is given. This makes
it possible to
deal with sections  without directly using any infinite
dimensional concepts. The
connection on the principal bundle defines a covariant
derivative on these
sections, which is found in finite form. In section 5  it is
shown that the
wavefunctions of the gauge--matter system are sections of an
associated vector
bundle. This is a direct consequence of Gauss' law at the
quantum level. Also a
general expression for the hamiltonian of the coupled system
is derived. In
particular, it shown that the Coulomb interaction arises from
the vertical part  of the gauge hamiltonian. This is not
immediately obvious from the purely
geometrical
picture. In section 6 we solve explicitly the
case of gauge theory
coupled to a single point particle. The solution  is
remarkably simple and
exhibits the phenomenon of confinement: the energy of the
ground state diverges
as the radius of the cylinder goes to infinity. This problem
has many features
in common with the charge--monopole system: it deals  with
vector bundles over
$S^3$ that are  the analogues of the line bundles  over $S^2$.
The
intermediate steps in our formulation are not manifestly
translation invariant,
as  a preferred point has to be picked in the definition of
the gauge group. (
This technical point is described in greater detail in the
text.) It is an
important check on our theory that the final results are
manifestly translation
invariant. In section 7 we  solve the case of two particles (
one in the fundamental and  the other in the  conjugate--
fundamental representation) coupled to the gauge field. This
is  a model
of a meson in a cylindrical space--time. It  turns out that
the obvious choice
of the meson wavefunctions ( separately periodic in the
positions of the two
particles)  leads to technical difficulties and a physically
unreasonable
answer. More consistent and physical answers are obtained by
requiring
periodicity only in the center of mass variable.  There could
be analogous
subtleties in  many body and  field theory problems.

\sect{ Review of Principal Fiber Bundles}

	We shall first define the notion of a Principal Fiber
Bundle \[kobayashi],\[isham].
Let $P$ be a manifold  $G$  a Lie group acting on $P$. The  action of $G$ on
$P$
 is defined by a map $R:G\times P\to P$
\beq
R (g,p) \mapsto p g^{-1},
\eeq
where $p \in P$ and $g \in G$. We will often write $R_{g^{-1}} p=pg^{-1}$.
This action is said to be  free if
$$p g^{-1} = p \Rightarrow g = e$$
where $e$ is the identity element of $G$.

	Points on $P$ connected by this action to the point $p \in P$ define an
equivalence class called the fiber containing $p$. We require that the
resulting quotient space $M \equiv P/G$ is itself a manifold with the quotient
topology. Let $\pi$ denote the projection map from $P$ to $M$. Let for every
open cover $\{ U_{\alpha} \}$
of M the inverse image $\pi^{-1}(U_{\alpha})$  is diffeomorphic to
$\{ U_{\alpha} \} \times G$, with the diffeomorphism being given by
$$\psi_{\al} \colon \pi ^{-1}(U_\alpha) \mapsto U_\alpha \times G,
\psi_{\al} (p)=(\pi (p),\phi_{\al} (p))$$
where
$$\phi_{\al} \colon {U_{\al}} \mapsto G, \phi_{\al}
(pg^{-1}) = \phi_{\al} (p)g^{-1}.$$
Such a fiber preserving diffeomorphism is called a local trivialization.
If such a local trivialization exists for every element of the open cover
$\{ U_{\alpha} \}$, then the set $\{ P, \pi, M, G\}$ is called a Principal
Fiber Bundle (PFB) and is denoted by $G \mapsto P \mapsto M$. $P$ and $M$ are
 called the total and base space
respectively.

	A local section of a PFB is a differentiable map
$s_\alpha \colon U_\alpha \mapsto \pi ^{-1} (U_\alpha)$ with the property
$\pi \circ s_\alpha = i_M$. Local sections can be seen to be in one to one
correspondence with local trivializations of a PFB by taking
$$\psi_{\al}(x, g) = s_{\al}(x) g^{-1}, \space x \in M, g\in G.$$
This also shows that if a PFB admits a global section then it is diffeomorphic
to $M \times G$. Such PFB's are said to be trivial.

	Next we introduce the concept of an Associated  Bundle .
Let $N$ be a  manifold
carrying an action $\rho$ of $G$ and let $C = \{(p,n)\}, p \in P ,
n \in N$. On $C$ define an equivalence relation $\sim$ :
$$(p,n) \sim (p g^{-1}, \rho(g) n).$$
Let $E = C/{\sim}$ and let us define a projection map
$\widehat \pi \colon E \mapsto M$ given by $\widehat \pi (p,n) = \pi (p)$. If
we require that for any open cover $\{U_{\al}\}$ of $M$ each image
$\widehat \pi ^{-1} (U_\alpha)$
be an open submanifold of $P \times N/G$, then the resulting structure is
called an Associated Bundle and is denoted by $N \mapsto E \mapsto M$.

	It can be shown that every
associated bundle admits a global section. Let the set of
global sections of an
associated bundle be denoted by $\Gamma(E)$. We can then show that
there is an isomorphism between $\Gamma(E)$ and the
set of equivariant maps $\Omega(P)$ from $P$ to $N$, where
$$\Omega(P) = \{ f:P \mapsto N | f(pg^{-1})=\rho (g) f(p) \}.$$

Two special cases of associated bundles are of particular relevance to us.
If $N$ happens to be a vector space $V$ and $\rho$ a representation of $G$ on
$V$, then the corresponding associated bundle is called an Associated Vector
Bundle (AVB). They are useful for describing matter fields. If on the other
hand $N$ happens to be the group $G$ and $\rho$ is the adjoint action of
$G$ on itself, then the corresponding associated bundle is called the
automorphism bundle. They are relevant for describing gauge transformations, to
which we turn next.

	Given a PFB, we can define the concept of a gauge transformation.
A gauge transformation is a fiber preserving diffeomorphism
$h \colon P \mapsto P$ such that $h(p g^{-1}) = h(p) g^{-1}$.
The set of such transformations denoted by ${\cal G}$ is
isomorphic to the set of equivariant maps from $P \mapsto G$, given by,
\beq
\Omega (P)_{adj} = \{ f:P \mapsto G | f(pg^{-1})=gf(p)g^{-1}\}.
\eeq
The isomorphism is then given by a map
\beq
\Phi : \Omega_{adj} \mapsto {\cal G}(P), \Phi(f)(p) = p f(p).
\eeq
By our earlier discussion ${\cal G}$ is also
isomorphic to the set of sections of the automorphism
bundle. However, if either the PFB or the automorphism bundle is trivial then
the gauge transformation reduce to maps from $M \mapsto G$. The set of gauge
transformations form a group called the gauge group.

	The construction outlined so far has been primarily topological. We now
turn to the description of geometric properties of the structures introduced
before. First, the group action (1) on $P$ provides a natural isomorphism, by
pushforward, between the Lie algebra ${\underline G}$ of $G$ and a subspace
$V_p$ of the tangent space $T_p $ at $p$. $V_p$ is called the vertical subspace
at $p$, and an element of $V_p$ is given by
\beq
V_\lambda= { d \over dt} (p e^{- \lambda t}) \bigm|_{t=0}, \space \lambda \in
{\underline G}.
\eeq
$V_\lambda$ is called a vertical vector  at point $p \in P$.

	At any given point $p$ the tangent space $T_p$ can be decomposed into
the vertical and a horizontal subspace $H_p$ such that
$ T_p = V_p + H_p$. However, unlike in the case of the vertical subspace, there
is no unique natural definition of the horizontal subspace. We next turn to the
concept of a connection which assigns a unique horizontal subspace at each
point $p$.

A smooth assignment of the horizontal subspace $H_p$ on all of $P$
is called a connection if :

 \item{1.} $V_p \oplus H_p = T_p$
 \item{2.} $H_{pg^{-1}}={(R_{g^{-1}})}_{*} H_p$
where ${(R_{g^{-1}})}_{*}$ is the pushforward induced by (1).

	Another equivalent way to specify a connection is given by
a Lie algebra valued one form $A \in \Lambda^1 \otimes \underline G$
(called the connection one form or simply the connection) satisfying

 \item{1.} $ i_{V_\lambda}(A)= -\lambda$ for any vertical vector
 \item{2.} $( R_{g^{-1}})_*(A) = ad_g(A) $ for any $g \in G$.

$H_p$ is then defined to be the kernel of $A_p$.
We mention without proof that every PFB admits a connection.

In a similar vein, we next define the space of equivariant $k$-forms, denoted
as
$(\Lambda^k (P) \otimes W)_{eq}$. These are $k$-forms in $P$
valued in the vector space $W$ which carries a representation $\rho$ of $G$ and
satisfy the relations
 \item{1.} $i_{V_\lambda}\omega=0$ for all $V_{\lambda} \in V_p$
 \item{2.} $ (R_{g^{-1}})^* \omega_{pg^{-1}}= \rho (g^{-1}) \omega_p$

The specification of a connection can be used to define a {\it covariant
derivative } on the PFB;
$$ D_A \omega : \Lambda^k (P)\otimes W \mapsto \Lambda^{k+1} (P)\otimes W$$
$$(D_A\omega)(u_0,u_1,...,u_k)=d\omega(u^H_0,u^H_1,...,u^H_k)$$
where $u_i=u^H_i+u^V_i$ gives the decomposition of a vector into horizontal
and vertical parts respectively.

The covariant derivative takes a particularly simple form if we consider
equivariant forms:
$$ D_A \omega : (\Lambda^k (P)\otimes W)_{eq}
\mapsto (\Lambda^{k+1} (P)\otimes W)_{eq}$$
$$ D_A \omega=d\omega + \underline \rho(A) \wedge \omega$$
for $\omega \in (\Lambda^k (P) \otimes W)_{eq}$
, where $\underline \rho$ is the representation of the Lie algebra
$\underline G$ induced by $\rho$ on W. The rhs of the last equation
also belongs to the space
of equivariant forms. This means that we can take a covariant derivative
again in the same way, getting
$$D_A^2 \omega= \underline \rho(dA + {1 \over 2} [A,A]) \wedge \omega$$
where $F \equiv dA + {1 \over 2}[A,A]$ is called the {\it curvature
two-form} defined by the connection $A$. It is possible to show that
$F= D_AA$.

	Having defined the connection we next study its behaviour under gauge
transformations. For this recall the definition of gauge transformations given
in (2) and (3). The transformation of $A$ is defined in
terms of the pullback $\Phi(f)^{- 1 *}$ induced by $\Phi(f)$ and is given by
\beq
A^{f} \equiv \Phi(f)^{-1*} A=fAf^{-1} + fdf^{-1}
\eeq
If the principal bundle is trivial, we can think of the connection simply as a
 Lie algebra--valued 1--form on $M$. This 1--form is obtained by pullback with
a
 global section of the principal bundle. A gauge transformation can then be
 thought of as a function $g:M\to G$. It describes a change of the
 trivialization ( global section); its effect on the connection 1--form is
\beq
	A\to gAg^{-1}+gdg^{-1}.
\eeq
The case of interest to us will in fact be of  this type.

In the quantum theory, we are interested in the space of all connections on a
 given principal bundle. Two connections are physically equivalent if they are
 related  by a gauge transformation. It is a remarkable fact that the space of
 connections modulo gauge transformations is itself an infinite dimensional
 principal bundle. This bundle can be  nontrivial even when  the original (
 finite dimensional) principal bundle is trivial.  All the geometric ideas
 discussed above can be applied to this infinite dimensional case, which will
be
 the way to get a nonperturbative formulation of {\it quantum} Yang--Mills
 theories \[singer].

	Let $\cal A$ denote the space of all connections on the PFB $P$.
${\cal A}$ and ${\cal G}$ are both infinite dimensional spaces, with ${\cal A}$
carrying an action of ${\cal G}$. We can therefore attempt to construct a PFB
${\cal G} \mapsto {\cal A} \mapsto {\cal A}/{\cal G}$. However the action of
${\cal G}$ on ${\cal A}$ may  not be free . For example if  $P$ admits a flat
 connection, the point $A=0 \in {\cal
A}$ is left invariant by the constant gauge transformation.
Demanding that an infinitesimal gauge action be free implies that
\beq
A'-A=t[A,\Lambda ]+td\Lambda =0 \quad {\rm iff} \quad \Lambda =0,
\eeq
where $\Lambda \colon P \mapsto {\underline G} \in \underline {\cal G}$
($\underline {\cal G}$ denotes the Lie algebra of the group ${\cal G}$.)
The above requirement can be satisfied in two different ways :
\item {i.} Restrict connections to satisfy this property, that is
$A \in {\cal A}^{irr}$, the space of irreducible connections.
\item {ii.} Restrict ${\cal G}$ to ${\cal G}_0= \{ f \in {\cal G}|
f(p_0)=e, \quad {\rm identity\ of} G \} $ for all $p_0 \in \pi^{-1}(x_0)$
where $x_0$ is an arbitrary point on $M$.
${\cal G}_0$ is called the space of pointed gauge transformations.
	We choose here to follow the second method (see next section for
explanation) and consider the PFB ${\cal G}_0 \mapsto {\cal A}
\mapsto {\cal G}_0/{\cal A}$.

It is now possible to show that the principal bundle $\cG_0\to \cA\to
 {\cA/\cG_0}$ admits a natural connection.
	Following the earlier discussion in the case of finite dimensional PFB,
we introduce the concepts of vertical and horizontal vectors. The vertical
subspace consists of vectors of the type
\beq
V_\Lambda={d \over dt}(e^{-t\Lambda}Ae^{t\Lambda} +e^{-t\Lambda}d\Lambda)
|_{t=0} =d\Lambda+[A,\Lambda]
\eeq
where $\Lambda \in \underline {\cal G}_0: P \mapsto \underline G$.

	We next assume that the original base space $M$ has a Riemann metric defined
on
it. This induces an inner product  on the tangent space $T {\cal A}$, which is
the space of equivariant one forms $(\Lambda^1(P)
\otimes \underline G)_{eq}$. This
metric is given by
\beq
d^2(A_1,A_2)=\| A_1-A_2\|^2=\int_M < A_1-A_2,*(A_1-A_2)>
\eeq
where "*" is the Hodge star operator defined by the metric on $M$ and $<.>$
denotes the trace operation on the Lie algebra ${\underline G}$. This metric
has
the property that $d(A_1^g,A_2^g)=d(A_1,A_2)$, where $A_i^g, i=1,2$ are the
gauge transformed versions of $A_i$.

	We can now define the horizontal subspace ${\cal H}_A$
to be one which is orthogonal
to the vertical subspace with respect to this metric. Therefore,
${\cal H}_A= \{ \xi \in (\Lambda^1(P) \otimes \underline G)_{eq}$ such that
\beq
\int_M < *\xi , d\Lambda + [A,\Lambda] > =0 \quad
\forall \Lambda \in \underline {\cal G}_0  \}.
\eeq
Integrating by parts and using the fact that $\xi$ is an equivariant
one-form we may rewrite this expression as,
$$ D_A *\xi =0  \qquad {\rm iff} \quad \xi \in {\cal H}_A.$$
This equation need not be true at the point in the fiber of  $p_0$  since
 $\Lambda$ should vanish there. The solution is in general a distribution.

These ideas are developed in more detail in the next section, in the special
 case where $M=S^1$ and $P$ is the trivial $G$--bundle over it. Even in this
 case, the bundle $\cG_0\to \cA\to {\cA/\cG_0}$ will be nontrivial. The base
 manifold will be finite dimensional; in fact ${\cA/\cG_0}=G$ in this case. The
 ideas of associated vector bundle, connection 1--form, curvature etc.
 introduced in this section will be useful in formulating and solving the
 quantum theory of Yang--Mills fields coupled to point particles. Some of
these
 techniques can be generalized to higher dimensions, \[singer]  although
we no longer expect the problem to be exactly solvable.

\sect{ The Bundle  of Connections}

Let space--time be a cylinder $S^1\times R$ and $G$ a
connected, simply connected compact
Lie group. In this case, a principal bundle with structure
group $G$ over space  $S^1 $ will be equivalent to the trivial
bundle. Then a connection on this principal bundle can be
thought of as a $\un{G}$--valued 1--form $Adx$ on $S^1$. We will use a
coordinate $x$  on $S^1$, with $0\leq x\leq 2\pi$.
This 1--form is the Yang--Mills potential  or `gauge field'.
Since the form has only one component, the   gauge field  can  also be thought
of as a
function
$A:S^1\to \un{G}$. Most of the time we will
use this simplification. A gauge transformation
 $g:S^1\to G$ acts on this as follows:
\beq
        A\mapsto gAg^{-1}+g\d g^{-1}/dx
.\eeq
Gauge fields that differ only by a gauge transformation are to
be regarded as
physically identical.

 Let $\cA=S^1\un{G}$ be the space of gauge potentials and
$\cG=S^1G$ the group
of gauge transformations. All gauge  transformations
are in the connected component of
identity in $S^1G$, since  $\pi_0(S^1G)=\pi_1(G)=0$.
Ideally we would like to formulate gauge theory directly on
the quotient space
$\cA/\cG$, so that it will be manifestly gauge invariant. But
there is an
immediate problem with this strategy. The action of the
 group $\cG$ has fixed points, so that the quotient is not a
smooth manifold.
The fixed points arise whenever there is a non-trivial solution
to the equation
\beq
        \D g+[A,g]=0;
\eeq
i.e., whenever the connection $A$ is reducible. These singular
points in $\cA$
can lead to unwieldy boundary conditions on the wavefunctions.

 There are two
ways around this difficulty:\hfill\break
1. consider  the quotient $\cA^{\hbox{irr}}/\cG$, of the space
of irreducible
connections, or,\hfill\break
2. consider the quotient $\cA/\cG_0$ by the subgroup of
`based' gauge
transformations, $\cG_0=\{g\in \cG|g(0)=1\}$. $\cG_0$ acts without
fixed points on $\cA$. This leaves
the constant part ( often called `global' part, $\cG/\cG_0=G$)
of the gauge group . This can be taken care of at the end by
requiring
the wave functions to be equivariant under $G$.

We will follow the second method, as it seems to have a closer
relation to the
usual perturbative formulation of gauge theories. ( The flat
connection, which
is the starting point for perturbation theory, is excised in
the first
approach). But it has the disadvantage that translation
invariance is not
manifest, as a special point $0$ is picked out in the
definition of $\cG_0$. We will see that in the end,
translation invariance is recovered.

There is then a principal fiber bundle $\cG_0\to \cA\to
\cA/\cG_0$.
The only gauge invariant information contained in $A$ on a
circle is its Wilson
loop. Therefore $\cA/\cG_0$ is just $G$ and the projection map
of the bundle is
the Wilson loop.
    (It is a special property of $1+1$-dimensional
gauge theory that this quotient is finite dimensional.)

More explicitly, define the parallel transport operator
$\t{W}:\cA\times S^1\to
G$
\beq
        {d \t{W}(A,x)\over dx}+A\t{W}(A,x)=0\quad \t{W}(A,0)=1
.\eeq
Under a gauge transformation,  $A\to gAg^{-1}+gdg^{-1}$,
$\t{W}(A,x)\to g(x)\t{W}(A,x)g(0)^{-1}$.
 The  Wilson loop is defined to be $W(A)=\t{W}(A,2\pi)$. This
is clearly invariant under the
action of $\cG_0$. Under the action of a constant gauge
transformation,
$W(A)\mapsto gW(A)g^{-1}$. Since the curvature of all
connections $A$ are zero, $W$ determines $A$ upto an action of
$\cG_0$. Thus the map $W:\cA\to G$ is indeed the projection
map of a principal fiber bundle. Once can construct local
sections and transition functions for this principal bundle
and verify that it is nontrivial.

Let us digress a little to consider  the special case
$G=SU(2)$. As a manifold, the base $SU(2)=S^3$. Now,
principal bundles on $S^3$ are labelled by $\pi_2$ of the
structure group. Hence there are no such nontrivial bundles
with a compact  Lie group. ( $\pi_2$ is zero in this
case). Yet, if the structure group is the based loop group
$S^1SU(2)_0$, there are nontrivial bundles over $S^3$:
\beq
     \pi_2(\cG_0)=\pi_3(G)=Z.
\eeq
The bundle we get over $S^3$ is the one corresponding to the
fundamental generator of $\pi_3(G)$. Thus the bundle we are
studying is in many ways the analogue of the bundle $U(1)\to
SU(2)\to S^2$, which arises in the study of magnetic
monopoles\[monopole]. We will see that the wavefunctions of our
physical system are  sections of an associated vector bundle,
analogous to the `monopole harmonics'.

Another interesting feature of our principal bundle $\cG_0\to
\cA\to G$ is that the total space is contractible. Therefore,
this is a model for the `Universal bundle'  \[milnor] of the gauge group
$\cG_0$. This feature is independent of the fact that
physical space is $S^1$. However it is unusual for the base
manifold of a universal bundle ( called the classifying space)
to be  finite dimensional. ( The precise statement is that the
classifying space is homotopic to $G$). This means that a
principal bundle over any manifold $M$, with structure group
$\cG_0$ is the pullback of our bundle $\cG_0\to\cA\to G$ through a map
$\phi:M\to G$. This $\cG_0$--bundle on $M$ is determined by
the homotopy class of the map $\phi$. More generally, this suggests that the
total space of any universal bundle can be thought as a space of connections.
This point of view might be useful to construct gauge theories with finite
dimensional or even discrete gauge groups. We will not pursue this idea here.

Vectors in $\cA$ can be thought of as functions $\xi:S^1\to
\un{G}$.
A vertical vector field in $\cA$ is one that points along the
fiber:
\beq
        \xi=\D\lambda+[A,\lambda],
\eeq
where $\lambda$ is a map from $S^1 \to \underline G$.
A connection on $\cA$ is a splitting of the tangent space into
a
vertical and a `horizontal' piece. We can construct a
connection on $\cA$,
starting from
the  obvious gauge invariant  Riemannian metric on $\cA$. The distance
$d(A_1,A_2)$ between any two points in $\cA$ is given by,
\beq
        d^2(A_1,A_2)=-\int {dx\over 2\pi}\tr(A_1-A_2)^2
.\eeq
A vector field
$\eta_h(A,x)$ is defined to be horizontal if it is orthogonal
to any vertical
vector:
\beq
        \int \tr \eta_h(A,x)(\D \lambda+[A(x),\lambda(x)])
dx=0,
\quad\hbox{for all $\lambda$ such that }\;\;
\lambda(0)=0
.\eeq
     That is,
\beq
        \D\eta_h+[A(x),\eta_h(x)]=k\delta(x)\;\;\hbox{for some
constant } k
.\eeq\label{one}
The delta function on the r.h.s. tells us that horizontal vectors $\eta_h$ may
have a discontinuity at the point $x=0$.

Thus a horizontal vector field is a function $\eta_h:\cA\times
S^1\to \un{G}$,
\beq
        \eta_h(A,x)=
          {1\over \surd (2\pi)}\t{W}(A,x)\eta\t{W}(A,x)^{-1}
.\eeq
( The constant ${1\over \surd (2\pi)}$ is put in for later
convenience.)
The function
 $\eta_h(x)$ is discontinuous at $x=0$ ( or $2\pi$) because of
the delta
function
in \(one).

It is useful to calculate the infinitesimal change in
$\t{W}$,
 $\t{W}\mapsto \t{W}+t\t{w}+O(t^2)$, when $A\mapsto
A+{t\over \surd(2\pi)}\t{W}(x)\eta\t{W}(x)^{-1}$. We have,
\beq
\D\t{w}+A\t{w}+{1\over \surd (2\pi)}\t{W}(x)\eta\t{W}(x)^{-1}
\t{W}=0
\eeq
so that
\beq
        \t{w}(A,x)=-{x\over \surd(2\pi)} \t{W}(A,x)\eta
.\eeq
 Also, the change in $W(A)$ is
\beq
W(A)\mapsto W(A)-t\surd(2\pi) W(A)\eta+O(t^2)
\eeq

Thus if $\eta\in \un{G}$ is independent of $A$ as well as $x$,
 the above horizontal vector
field descends to a left--invariant vector field on $G$.
Conversely,  the
left invariant vector field, $\eta\in \un{G}$ can be lifted to
a horizontal
vector $ \eta_h(A,x)={1\over
\surd(2\pi)}\t{W}(A,x)\eta\t{W}(A,x)^{-1}$ in $\cA$.

\BEGINIGNORE{
The connection can also be described by   a one--form $\omega$
on $\cA$ valued
 in
$S^1\un{G}_0$. $\omega$ must give zero on a horizontal vector
and $\lambda$ on a vertical  vector $\D \lambda +[A,\lambda]$.
Thus $\omega$ is given  at the point $A\in
\cA$  by,
\beq
       i_{\xi} \omega_{A}=\Delta_A^{-1}(\D\xi+[A,\xi])
\eeq
where $\Delta_A$ is the covariant Laplacian.( The equation
$\Delta_A\omega=\rho$ has a
solution only if the source $\rho$ is orthogonal to the kernel
of $\Delta_A$.
But the kernel of $\Delta_A=d_A^2$ is the same as that of
$d_A=\D+\ad A$;i.e.,
covariantly constant functions.  $\rho=d_A \xi$  is indeed
orthogonal to such
 functions. Furthermore, the solution is unique once it is
required to vanish at $0$.)
}ENDIGNORE

\sect{ Associated vector bundles}

Let ${\cal F}$ be a vector space carrying a  representation
${\cal \rho}$  of $
\cG_0$.
Then there is an associated vector bundle
${\cal F}\to \cA\times {\cal F}/\cG_0\to G$. A section of this
vector bundle is
a function $\psi:\cA\to {\cal F}$ such that
\beq
        \psi(gAg^{-1}+g\D g^{-1})={\cal \rho}(g)\psi(A),
\;\;\hbox{if}\; g(0)=1
.\eeq
We will see that the wavefunctions of \ym theory coupled to matter satisfy this
 `equivariance condition'. Hence they are naturally identified with sections of
an associated bundle.

 There is a representation of $S^1G$ on $S^1V$, when $V$
carries  a finite dimensional
representation ( also   called $\rho$) of $G$, given by pointwise
multiplication.
Let ${\cal V}$ be the vector bundle associated to this
representation. It will
be important to find a simple description of the sections of
${\cal V}$, as
they describe one particle wavefunctions. The ideas can then be
easily generalized to multiparticle wavefunctions.

 A function $\psi:\cA\to S^1V$ is the same as a function
$\psi:\cA\times S^1\to
V$. In order to be a section of ${\cal V}$, it  must satisfy
\beq
        \psi(gAg^{-1}+g\D g^{-
1},x)=\rho(g(x))\psi(A,x)\quad\hbox{for all }\; g\in
\cG_0
.\eeq\label{sectionone}
This is however a complicated way of describing a section: it
involves functions of an infinite number of variables. We will
now  find an
alternative description in terms of functions of a finite
number of variables,
which is more useful for later calculations.

Solutions to \(sectionone) are of the form
\beq
\psi(A,x)=\rho\big(\t{W}(A,x)\big)\phi(W(A),x)
.\eeq
This gives a second, much simpler
characterization.\hfill\break
{\bf Defn.}A section of ${\cal V}$ is  a function
 $\phi:G\times [0,2\pi]\to V$ satisfying  the quasi--
periodicity condition
\beq
        \phi(W,0)=\rho(W)\phi(W,2\pi)
.\eeq\label{sectiontwo}
We may use this quasi--periodicity to extend $\phi$ to a
function on the whole
real line:
\beq
        \phi(W,x+2\pi )=\rho(W^{-1})\phi(W,x) ,
.\eeq

Since our principal bundle has a natural connection, given a left invariant
vector field $\eta\in \un{G}$, there
must be a covariant
derivative operator $\nabla_{\eta}$ on  such sections. We can
show that this is
 simply
\beq
\nabla_\eta\phi(W,x)=\surd(2\pi)[\lie_{\eta}-{x\over
2\pi}\rho(\eta)]\phi(W,x)
,\eeq
$\lie_{\eta}$ being the Lie derivative w.r.t. left invariant
vector fields:
\beq
\lie_{\eta}\phi(W,x)=\lim_{t\to 0}{\phi(W(1- t\eta),x)-
\phi(W,x)\over t}
.\eeq

To get the above formula for $\nabla_\eta$, recall that the covariant
derivative of a
section $\psi:\cA\to
S^1V$ is just the Lie derivative along a horizontal vector field on
$\cA$. Given $\eta\in \un{G}$, there is a corresponding
horizontal vector field
 in $\cA$, $ \eta_h(A,x)={1\over
\surd(2\pi)}\t{W}(A,x)\eta\t{W}(A,x)^{-1}$.
 The Lie derivative
of $\psi$ along this horizontal vector field is
\beq
  \lie_{\eta_h}\psi(A,x)=       \lim_{t\to 0}
{\t{W}(A+t\eta_h,x)\phi(
W(A+t\eta_h),x)-\t{W}(A,x)\phi(A,x)\over t}
.\eeq
We can calculate the r.h.s explicitly and then extract the
covariant derivative in terms of $\phi$:
\beq
        \nabla_{\eta}\phi(W,x)=\rho\big(\t{W}(A,x)^{-
1}\big)\lie_{\eta_h}\psi(A,x)
.\eeq

We pause to  make a technical remark. The space $\cA$ has been implicitly
 assumed to be consisting smooth  functions $A:S^1\to \un{G}$. This means that
a
 one--form in $\cA$ would be a distribution on $S^1$ valued in $\un{G}$. In
 particular it could be a discontinuous function on $S^1$. The connection on
the
 bundle $\cG_0\to \cA\to G$ can be thought of  as a one--form on $\cA$. It
turns
 out that it  has a discontinuity  as a function on $S^1$. This also means that
 the covariant derivative of a quasi--periodic function is no longer
 quasi--periodic. This is not an inconsistency.

Given  $\eta_1,\eta_2\in \un{G}$ the curvature of this
connection can be
calculated to be
\beq
        \Omega_{\eta_1,\eta_2}(x)=
[\nabla_{\eta_1},\nabla_{\eta_2}]-\nabla_{[\eta_1,\eta_2]}=
({x\over \surd(2\pi)}-1){x\over \surd(2\pi)} \rho ( [\eta_1,\eta_2])
.\eeq
Viewed as a function on $S^1$, this is discontinuous;  that is allowed for the
 technical reason mentioned above.

Under a gauge transformation $g\in \cG$ ( possibly with
$g(0)\neq 1$), the
section $\phi(W,x)$ transforms as follows:
\beq
        \phi(W,x)\mapsto g(0)\phi(g(0)^{-1}Wg(0))
.\eeq
It is straightforward to check that this is consistent with
the
quasi--periodicity condition.

The operator $d_A:\psi(A,x)\mapsto \D\psi(A,x)+A(x)\psi(x)$
maps  a section
$\psi:\cA\times S^1\to V$
 of
the associated bundle to another  section.
If we think of sections instead as functions $\phi(W,x)$
satisfying the quasi--periodicity condition, this operator
becomes just $\D$. This is part of the simplicity of thinking in terms of
 $\phi$ rather than $\psi$.
We can also describe the one--particle wave--functions of the
systems in the
momentum basis, as functions $\t{\phi}:G\times R\to V$, with
\beq
        e^{2\pi ik}\t{\phi}(W,k)=W^{-1}\t{\phi}(W,k)
\eeq \label{sectionthree}
in which case the covariant derivative will be
\beq
        \nabla_{\eta}=\surd(2\pi)[\lie_{\eta}-{\rho(\eta)\over
2\pi i}\d/dk]
.\eeq

\sect{ The Quantization of Gauge Fields Coupled to Matter}

The action for \ym theory coupled to matter can be  chosen to
be
\beq
        S=\int \tr E(\pdr_tA-\pdr_xA_0+[A_0,A])dxdt+\half\int
\tr
E^2dxdt-i\int\tr \rho(x)A_0(x) dx+S_m(A,\chi)
,\eeq
 $S_m$ being the action for the matter variables $\chi$
and $\rho(x)$ the charge density of
matter. The configuration space of the matter  will be called
 $Q$.
Classically, the variation with respect to $A_0$  leads to the
constraint equation ( Gauss' law):
\beq
        \D E+[A,E]-i\rho(x)=0.
\eeq
In fact $A_0$ is just the Lagrange multiplier that enforces
these constraints.
The constraints are   first class  in the sense  of
Dirac.
The action also implies that $A$ and $E$ are canonically conjugate
to each other.
Furthermore, we can  read off the hamiltonian
\beq
        H=-\int dx \tr E^2+H_m(A,\chi)
.\eeq
Here $H_m$ is the  hamiltonian of the matter fields.

 In the quantum theory, wave functions are functions
$\psi:\cA\times Q\to C$. Denote by $\cF$ the space of functions on $Q$ ( the
space of matter wavefunctions). Then, $\psi$ can be thought of as a function on
$\cA$ valued in $\cF$, $\psi:\cA\to \cF$.
 The electric field, being canonically conjugate to $A$, is
represented by the
 operator $E(x)={\de\over i\de A(x)}$. More precisely, upon
smoothing out with the ( Lie algebra  valued) function $\xi$,
\beq
     \int dx \tr \xi(x)E(x)\psi(A,\chi)=-
i\lie_{\xi}\psi(A)=-i\lim_{t\to 0}{\psi(A+t\xi)-
\psi(A)\over t}.
\eeq

 Also, upon quantizing
the matter field, the charge density $\rho(x)$ is given by an
operator. This operator must provide a representation of the  Lie algebra of
the gauge group on the Hilbert space of matter wavefunctions $\cF$.
That is, if$\rho(\lambda)=\int dx \tr
\lambda(x)\rho(x)$,
\beq
     [\rho(\lambda),\rho(\lambda')]=\rho([\lambda,\lambda']).
\eeq
Being first class, the  constraint equation becomes a
 differential equation  on the allowed ( `physical') wave--
functions. If $\lambda\in S^1\un{G}$, we can write it as
\beq
     -\int dx \tr d_A\lambda\; E\psi(A,\chi)=i\int \tr
\rho(x)\lambda(x)\psi(A,\chi).
\eeq
Thus, under an infinitesimal gauge transformation,
\beq
     \lie_{d_A\lambda}\psi=\rho(\lambda)\psi.
\eeq
Since our gauge group $\cG$ is connected, this may be integrated
to a constraint under finite gauge transformations:
\beq
     \psi(gAg^{-1}+gdg^{-1})=\rho(g)\psi(A).
\eeq
( To avoid proliferation of symbols, we use $\rho$ to denote
the representation of the group as well as the Lie algebra.)
If we restrict to gauge transformations that satisfy $g(0)=1$,
this is just the `equivariance' condition that $\psi$ be a
section of the associated vector bundle $\cA\times \cF/{\cG}_{0}$. They have to
satisfy in addition an equivariance condition under the `global' (
constant) gauge transformations $\cG/\cG_0=G$.
\beq
	\psi(gAg^{-1})=\rho(g)\psi(A)
\eeq

More generally, given any  vector bundle  $\cF\to {\cal T}\to G$
associated to the
principal bundle $\cG_0\to \cA\to G$, we have a theory of
matter coupled to \ym theory. Any representation of the gauge
group will provide such an associated bundle. \footnote{$^*$}{ There could be
additional restrictions for the theory to be physically
reasonable. For example, the hamiltonian should represented by
a self--adjoint operator that is bounded below.} Actually all
that is necessary is a 1--cocycle of the group. It is thus possible to have
more general `matter fields' whose transformation law is not just a
representation of $\cG_0$, but instead depends on $A$.
We will only study here the case of bundles arising from some
obvious representations.

We now need to understand the meaning of the hamiltonian in
this language. The matter part of the hamiltonian is an
operator on each fiber and requires no further comment. ( We
will work out some special cases later.) We will show now that
the term $\int dx \tr E^2$ is the sum of a covariant Laplacian
in $G$ and the Coulomb energy.The topologically nontrivial
part is of course in the covariant Laplacian\[mickbook].
It is interesting that in our view the Coulomb energy arises
from the `kinetic energy' $\int dx \tr E^2(x) $  of the \ym field.

We know that
\beq
        \int \tr \eta(x)E(x)\psi(A)dx=E_{\eta}\psi(A)=
-i\lim_{t\to 0}{\psi(A+t\eta)-\psi(A)\over t}
.\eeq
 If $\eta$ is  a horizontal vector in $\cA$, this is the same
as the covariant
derivative ( upto  a factor of $-i$):
\beq
     \int dx \tr \eta(x)E(x)\psi=-i\nabla_\eta\psi\;\hbox{if $\eta$ is
horizontal}.
\eeq
  On the other hand, if $\eta = d_A \lambda $ is  a vertical vector, then
the component along this vertical direction is given by the
constraint on the
wavefunctions:
\beq
        E_{d_A\lambda}\psi(A,\chi)=-i\int dx
\lambda(x)\rho(x)\psi(A,\chi)
.\eeq
Let $\eta_{a}$,be a complete set of horizontal vectors and
 $d_A\lambda_m$  a  complete set of vertical vectors,
orthonormal  with respect to the metric on $\cA$:
\beq
        \int \tr \eta_a(x)\eta_b(x)=-\de_{ab}\quad \int \tr
d_A\lambda_m(x)d_A\lambda_n(x)=-\de_{mn}
.\eeq
 Then we can define   the operator $\int \tr E^2(x)$ to be
\beq
        \int \tr E^2(x) dx=-\sum_a E_{\eta_a}^2-\sum_m
E_{d_A\lambda_m}^2
.\eeq
The functions $\lambda_m$ must vanish at the origin in order
to be in $\cG_0$. Furthermore, they might have discontinuous
first derivatives at that point.

 From the above discussion, we get
\beq
        \int \tr E(x)^2 dx=\sum_a\nabla_{\eta_a}^2+
                \sum_m\int dxdy
\tr[\rho(x)\lambda_m(x)]\tr[\rho(y)\lambda_m(y)]
.\eeq
Now it is important to recall that $\lambda_m$ are orthonormal
with respect to an inner product that depends on $A$. Thus
$\sum \lambda_m(x)\otimes \lambda_m(y)=G_A(x,y)$ also depends
on $A$. We can compute  $G_A$ by expressing $\lambda_m$ in
terms of the eigenfunctions of the operator $d_A^2$:
\beq
     d_A^2\mu_n=-a_n \mu_n,\quad \int \tr \mu_m(x)\mu_n(x)dx=-
\delta_{mn}\quad \sum_m\mu_m(x)\otimes \mu_m(y)=-\delta(x,y).
\eeq
The eigenvalues $a_m$  are positive. ( The zero eigenvalue exists only if there
is a constant eigenvector; but that is excluded by the  condition
$\lambda_m(0)=0$.)
Now we see that $\lambda_m={1\over \surd a_m}\mu_m$ for
$a_m\neq 0$. Also, $\sum_m\lambda_m(x)\otimes
\lambda(y)=G_A(x,y)$ is the Green's function of the one--
dimensional  Laplace operator: $d_A^2G_A(x,y)=\delta(x,y)$.( The boundary
condition $G_A(x,0)=G_A(0,y)=0=G_A(2\pi,y)=0=G_A(x,2\pi)=0$
has to be imposed on $G_A$).

We can in fact  find $G_A$ more explicitly. By its definition, $G_A(x,y)$ is a
matrix in the adjoint representation. That is, under a gauge transformation, it
transforms as
\beq
	G_{gAg^{-1}+gdg^{-1}}(x,y)=\hbox{Ad}(g(x))G_A(x,y)\hbox{Ad}(g(y))^{T}.
\eeq
Using the definition of $\t{W}$, it is possible to check that
\beq
	G_A(x,y)=\Ad(\t{W}(A,x))G_0(x,y)\Ad(\t{W}(A,y))^{T}
\eeq
where
\beq
	G_0(x,y)= \half |x-y| + {xy \over {2\pi}} - \half (x+y).
\eeq

Thus the total hamiltonian of the gauge--matter system can be
written as
\beq
        H=-\nabla^2 +\int \tr \rho(x)\rho(y)G_A(x,y) dxdy+H_m(A)
.\eeq
We see that matter couples to the \ym field in two different
ways: through the Coulomb term and the covariant derivative.
 If we had not used the proper geometric language, we might have
missed the first
term.

It is  useful to write the Coulomb energy more explicitly. If we introduce an
orthonormal  basis $\eta_a$ in $\un{G}$ labelled by $a=1\cdots \dim G$, we can
write
\beq
	\int dxdy\tr
\rho(x)G_A(x,y)\rho(y)=\int dxdy\rho^a(x)\rho^b(y)G_A^{ab}(x,y).
\eeq
The $\rho^a(x)$ are operators on matter wavefunctions satisfying
\beq
	[\rho^a(x),\rho^b(y)]=f^{abc}\rho^c(x)\delta(x,y).
\eeq

\sect{ Yang--Mills field coupled to a single particle}

Now we consider the simplest possible kind of matter: a nonrelativistic point
particle of mass $\mu$, coupled to the gauge field. The wavefunction of the
matter field can be thought of as a function $f:S^1\to V$, $V$ being  finite
dimensional and carrying a unitary irreducible representation $\rho$ of
$G$.\footnote{$^*$}{ Strictly speaking, the wavefunctions need to be periodic
only upto a phase, in $S^1$. Such $\theta$--parameters will be ignored largely,
as our present aim is to construct the simplest quantum theory, not the most
general one.
}
The wavefunctions of the \ym--matter system will be functions $\psi:\cA\times
S^1\to V$ such that
\beq
	\psi(gAg^{-1}+gdg^{-1},x)=\rho(g(x))\psi(A,x).
\eeq
We already know that these can be written as
\beq
	\psi(A,x)=\rho(\t{W}(A,x))\phi(W,x)
\eeq
where the $\phi:G\times R\to V$ satisfy the constraints
\beqs{
	\phi(W,x+2\pi)=\rho(W^{-1})\phi(W,x)\cr
	\phi(gWg^{-1},x)&=\rho(g)\phi(W,x)\; \hbox{for}\; g\in G.\cr
}\eeqs
The first one is the condition for $\phi$ to define a section of the associated
vector bundle with fiber $\cF$ over $G$. The second is the equivariance
condition under the `global' or constant part of the gauge transformations.

Now it is clear right away that not all representations $\rho$ are allowed. For
example, if $G=SU(N)$ and $V=C^N$, the second equation becomes
\beq
	\phi(gWg^{-1},x)=g\psi(W,x).
\eeq
The only solution is $\phi=0$. To see this, consider the special case $g=W$:
\beq
	\phi(W,x)=W\phi(W,x)
\eeq
 and recall that $W$ is an arbitrary element of $SU(N)$ ( which may not have
eigenvalue $1$.)

This has a simple meaning. The condition says that the wave--function must be
invariant under the constant transformations, if the transformations of  the
gauge field and matter are both taken into account. Now, the gauge field,
described by $W$, transforms under the adjoint representation. There is no way
to form a singlet by combining a power of the adjoint representation and the
fundamental one. We see that in order  to have a nontrivial solution to the
constraints, the matter representation must have the center of $G$ in its
kernel.
( For $SU(2)$, these are the integer spin representations; for $SU(3)$, these
are the representations of zero triality.)

Let us now return to studying the constraints.
The second equation with $g=W^{-1}$ implies that
$\phi(W,x)=\rho(W^{-1})\phi(W,x)$.
This means we can simplify the  first constraint:
\beqs{
	\phi(W,x+2\pi)&=\phi(W,x)\cr
	\phi(gWg^{-1},x)&=\rho(g)\phi(W,x).\cr
}\eeqs
Thus the wavefunctions are just equivariant periodic functions $\phi:G\times
S^1\to \un{G}$.

The results of the last section can be used to simplify the hamiltonian.
In general,
\beq
	H=-\nabla^2+\int dxdy\tr \rho(x)G(x,y)\rho(y)+H_m(A).
\eeq
In our case the matter hamiltonian is ${-1\over 2\mu}(\D+\hbox{ad A})^2$ when
acting on  $\psi$.  After changing variables to $\phi$, it is just
\beq
	H_m\phi=-{1\over 2\mu}{d^2\phi\over dx^2}.
\eeq
{}From earlier arguments,
\beq
	-\nabla^2\phi(W,x)=2\pi \sum_a[\lie_{\eta_a}-{x\over 2\pi}\rho(\eta_a)]^2\phi
\eeq
where $\eta$ are an orthonormal basis in $\un{G}$ and $\lie_\eta$ is the
corresponding left--invariant  vector field on $G$.

This leaves the Coulomb energy. The charge density operator $\rho(x)$ is
defined by ,
\beq
	\rho(\lambda)\phi(W,x)=\int \tr\lambda(x)\rho(x)
dx\phi(W,x)=\rho(\lambda(x))\phi(W,x)
\eeq
If we use the orthonormal basis $\eta_a$ in $\un{G}$, the charge density is the
operator
\beq
	\rho^a(y)\phi(W,x)=\rho( \eta_a) \delta(y-x)\phi(W,x).
\eeq
That is, charge density is concentrated at the position of the particle.
The Coulomb energy $V$ becomes,
\beq
	V\Phi(W,x)=\int \tr
\rho^a(y)\rho^b(z)G_A^{ab}(y,z)dydz\phi(W,x)=\rho(\eta_a)\rho(
\eta_b) G_A^{ab}(x,x)\phi(W,x).
\eeq
This just describes the self--energy of the particle. Now, in matrix notation,
\beq
	G_A(x,y)=\Ad(\t{W}(A,x))\Ad(\t{W}(A,y))^T G_0(x,y).
\eeq
The first two factors cancel each other when $x=y$, since the adjoint
representation is orthogonal. Thus
\beq
	G_A^{ab}(x,x)=\delta^{ab}G_0(x,x),
\eeq
and
\beq
	V\phi(W,x)=G_0(x,x)\rho( \eta_a)\rho( \eta_a)\phi(W,x)=
C_2(\rho) G_0(x,x)\phi(W,x)
\eeq
where $C_2(\rho)$ is the quadratic Casimir of the  representation $\rho$.

Thus we see that the hamiltonian becomes
\beq
	H\phi(W,x)=\{-2\pi\sum_a[\lie_{\eta_a}-{x\over
2\pi}\rho(\eta_a)]^2+C_2G_0(x,x)-{1\over 2\mu}{d^2\over dx^2}\}\phi(W,x).
\eeq
We must find the eigenvectors of this $H$ subject to the constraints above.

The first two terms in the hamiltonian are not individually translation
invariant; yet the sum is invariant on the subspace of functions satisfying the
constraints. So we should be able to eliminate all explicit $x$ dependence from
the hamiltonian, using the constraints. The infinitesimal form of the
constraint
\beq
	\phi(gWg^{-1},x)=\rho(g)\phi(W,x)
\eeq
is
\beq
	(L_a+R_a)\phi(W,x)=\rho(\eta_a)\phi(W,x).
\eeq
Here, $L_a=\lie_{\eta_a}$  are the left invariant vector fields and $R_a$ the
right invariant vector fields on $G$. Note also that the operators $L_a^2$ and
$R_a^2$ are the same on $G$, since they are just different ways of expressing
the Laplace operator on $G$.
Now we can simplify  the first two terms in the hamiltonian using
\beq
	L_a\rho(\eta_a)\phi=L_a(L_a+R_a)\phi=\half(L_a+R_a)^2\phi=\half C_2(\rho)\phi
\eeq
The third step uses $L_a^2=R_a^2$. The $x$ dependent terms now cancel out.
 Thus the hamiltonian simplifies to just
\beq
	H\phi(W,x)=-[2\pi L_a^2+{1\over 2\mu}{d^2\over dx^2}]\phi(W,x).
\eeq
The constraints
\beq
	\phi(W,x+2\pi)=\phi(W,x)\quad \phi(gWg^{-1},x)=\rho(g)\phi(W,x)
\eeq
are also quite simple to solve.

This is to be compared to the corresponding result for pure \ym theory:
\beq
	H\phi(W)=-2\pi L_a^2\phi(W)
\eeq
 and
\beq
	\phi(gWg^{-1})=\phi(W).
\eeq
This is just  the  special case where $\rho$ is the trivial representation and
the particle therefore decouples from the gauge field.
The eigenfunctions were character functions of the various irreducible
representations; the eigenvalues then are the corresponding quadratic Casimirs.

The parameters of our problem are the gauge coupling constant $\alpha$, the
radius $R$ of the circle and the mass $m$ of the particle. We can use units in
which $\hbar=1$. Since we deal with nonrelativistic particles, the velocity
of light never enters the theory:$c\neq 1$. Thus there are two dimensions, (
say) length and time. The dimensions of the physical quantities are
\beq
	[A]=L^{-1},[E]=1,\quad [H]=T^{-1},\quad [\alpha]=T^{-1}L^{-1},\quad
[m]=TL^{-2},\quad
\eeq
In the above we have used units with $\alpha=R=1$, and $\mu=m\alpha R^3$ is the
only dimensionless parameter of the theory.If we express the hamiltonian
explicitly in terms of the physical parameters, we get
\beq
	H\phi(W,x)= -[\alpha R L_a^2 +{1\over 2m }{d^2\over dx^2}]\phi(W,x).
\eeq

Let us consider the solution when the gauge group is $SU(2)$ and the
 representation $\rho$  of the matter field has  dimension $2j+1$. If $j$ is
 half an odd integer, there is no solution to the constraint
\beq
	\phi(gWg^{-1},x)=\rho(g)\phi(W,x)
\eeq
at all ( except $\phi=0$). This is seen by considering $g=-1$, and noting that
 $\rho(g)=-1$ for such even dimensional representations. If $j$ is an integer,
 there is exactly one solution for each half--integer or integer  $l$  such
that
 $2l\geq j$. To see this, note that there is a representation $R$ of
 $SU(2)\times SU(2)\times SU(2)$ on the space of functions $\phi:SU(2)\to
 C^{2j+1}$:
\beq
	\phi(W)\to \rho_j(g_1)\phi(g_2^{-1}W g_3).
\eeq
The solutions to our condition, $\phi(W)=\rho(g)\phi(g^{-1}Wg)$ are invariant
 under the diagonal $SU(2)\subset SU(2)\times SU(2)\times SU(2)$ subgroup. The
 representation $R$ of $SU(2)^3$ above can be expanded in terms of irreducible
 ones:
\beq
	R=\oplus_{l=0,\half,1,\cdots} (j,l,l)
\eeq
 and can be reduced to the following representation of $SU(2)\times SU(2)$ by
 combining the actions of the last two:
\beq
	\oplus_{l=0,\half,1,\cdots} [(j,2l)\oplus(j,2l-1)\oplus
 (j,2l-2)\cdots,\oplus(j,0)].
\eeq
In order that there be an invariant when the last two $SU(2)$ are combined,
 there should be a term in the square bracket of the form $(j,j)$. In that case
 there will be exactly one such invariant vector. Thus to each
 $l=0,\half,1,\cdots$
 satisfying $2l\geq j$ there is one solution to the constraint.

Let us call this solution $\psi_l(W)$. Then the  eigenfunctions of the
 hamiltonian are $\psi_l(W)e^{ipx}$ and the eigenvalues are given by
\beq
      H\psi_l(W)e^{ipx}=[\alpha R 2\pi l(l+1)+{p^2\over 2m }]\psi_le^{ipx}.
\eeq
Here $l={j\over 2}, {j\over 2}+1,\cdots$ and $p$ is an integer.

In particular, the ground state corresponds to $p=0$, $l={j\over 2}$.
Thus even the ground state has energy of order $\alpha R$.
 This situation has a simple physical meaning: to form a singlet under  the
 global symmetry, the particle has to `borrow' some color from the gluon
sector.
 This costs an amount of color proportional to $\alpha R$.  In the limit where
 the radius of the circle is send to infinity keeping $\alpha$ fixed, the
energy
 of any particle carrying color will diverge linearly. This is just the
 expression of confinement in our simple situation. There is also a close
 analogy with the energy levels of the charge--monopole system \[monopole].

\sect{Bound State of Two Particles}

We will now consider the case of two point particles on a circle interacting
 through the Yang--Mills field. If the two particles  combine to form a color
 non--singlet, the ground state energy of the system will be of order $R$. This
 is because the center of mass variable of the system by itself will behave
 like the example considered above. Some color would be borrowed from the gluon
 sector to form a singlet, but that would increase the energy of the gluon
 sector by an amount of order $R$. Thus in the limit of large $R$, the low
lying
 states of the system will be in the sector where the two particle bound state
 is in the color singlet state ( if that is possible). We will study in detail
 the case of two particles in complex conjugate fundamental representations  of
 $G=SU(N)$, which allows for both possibilities. We might think of this a
 particle (`quark') and its antiparticle, interacting through the Yang--Mills
 field. Generalization to other cases is possible, but we do
not expect any essential changes.

It would be reasonable to assume that  wavefunction of the two particle system
 ( excluding the Yang--Mills degrees of freedom ) transforms under an action of
 $\cG$ as
\beq
	\psi(x,y)\to g(x)\psi(x,y)g(y)^{-1}.
\eeq
At first it seems reasonable also that the wavefunction should be  separately
 periodic in the coordinates $x$ and $y$. However, that leads to a physically
 unreasonable answer: the ground state energy will be order $\alpha R$ when the
 radius $R$ of the cylinder  is large.   Therefore we will only impose
 periodicity under the
 simultaneous shift of $x$ and $y$ by $2\pi$ ( in units where $R=1)$. If we
 shift $x$ alone by $2\pi$ keeping $y$ fixed, the particle would have to pass
 through the position of the antiparticle. In one--dimensional situations, it
 is possible that discontinuities in the wavefunction arise as result of this.
 The  allowed boundary conditions (at the points where the positions of the two
 particles coincide) is related to the self--adjoint extension of the
 hamiltonian.  Thus we will impose
\beq
	\psi(x+2\pi,y+2\pi)=\psi(x,y).
\eeq
Additional boundary conditions will become clear only after we know the
 hamiltonian in explicit form.

 We can regard the matter  wavefunctions as functions
$\psi:(R\times R)/2\pi Z\to V\times V^*$, where the vector space $V$ carries
the
 fundamental representation  of $G$.  The wavefunctions of the matter--gauge
 system is then the space of sections of the corresponding associated vector
 bundle. These sections can be thought of as functions $\Psi:\cA\times
 \Big((R\times
 R)/2\pi Z\Big)\to V\otimes V^*$ satisfying the equivariance condition
\beq
	\Psi(gAg^{-1}+gdg^{-1},x,y)=g(x)\Psi(A,x,y)g(y)^{-1}.
\eeq
To be a section of the associated vector bundle, it is sufficient that this
 condition be satisfied for $g$ with $g(0)=1$. However, physical wavefunctions
 must in fact satisfy this for all $g$, even those that do not become the
identity
  at the
 origin.
We can solve this equation as before by the ansatz
\beq
	\Psi(A,x,y)=\tW(A,x)\Phi(W,x,y)\tW(A,y)^{-1}.
\eeq
Now the equivariance is automatic if $g(0)=1$.  $\Psi$ will be equivariant
under
 the full gauge group if $\Phi$ satisfies the constraint
\beq
	\Phi(gWg^{-1},x,y)=g\Phi(W,x,y)g^{-1}.
\eeq
Also, the periodicity condition becomes, in terms of $\Phi$,
\beq
	\Phi(W,x+2\pi,y+2\pi)=W^{-1}\Phi(W,x,y)W.
\eeq

The hamiltonian operator acting on $\Psi$ is,
\beq
	H\Psi=-\int \tr E^2(z)\Psi dz-{1\over 2m}[{\pdr\over \pdr x}+A(x)]^2\Psi
		-{1\over 2m}[{\pdr\over \pdr y}+A^*(y)]^2\Psi.
\eeq
We have assumed that the two particles have the same mass. There is no
essential
 difference if they are not equal.

 We can now decompose the first term into a
 horizontal and a vertical  part as before.
The derivative along the horizontal direction becomes,
\beq
	E_{\eta_a}\Psi(A,x,y)=i\surd(2\pi)\tW(x)[\lie_{\eta_a}-{x\over
 2\pi}\eta_{aL}+{y\over 2\pi}\eta_{aR}]\Phi\tW(y)^{-1}.
\eeq

 It is again possible to express the contribution of the vertical derivatives
 to the  Hamiltonian in terms of the Green's functions. It becomes the sum of
 the self--energies  and
 the interaction  energy of the particle and the antiparticle. In more detail,
\beq
	\sum_m E_{d_A\lambda_m}^2\Psi=\sum_m[
 -\lambda_m(x)\lambda_m(x)\Psi-\Psi\lambda_m^{\dag}(y)\lambda_m^{\dag}(y)-
2\lambda_m(x)\Psi\lambda_m(y)^{\dag}].
\eeq
We already know how to simplify the first two terms. The last term may
be simplified using the identity
\beq
	\sum_m \lambda_m(x)^i_k\lambda_m(y)^{\dag
 l}_j=G_0(x,y)[P^i_j(x,y)P^{*l}_k(x,y)-{1\over N}\delta^i_k\delta^l_m]
\eeq
where $P^i_j(x,y)=(\tW(x)\tW(y)^{-1})^i_j$. This term  describes the
interaction
 energy of
 the particle--antiparticle pair.

Now it is possible to write the hamiltonian in terms of $\Phi$:
\beqs{
	H\Phi&=-2\pi\sum_a[\lie_{\eta_a}-{x\over 2\pi}\eta_{aL}+{y\over
 2\pi}\eta_{aR}]^2\Phi+\cr
&C_2[G_0(x,x)+G_0(y,y)]\Phi+2G_0(x,y)[\tr\Phi-{1\over N}\Phi]-{1\over
 2m}{\pdr^2\over \pdr x^2}\Phi-{1\over 2m}{\pdr^2\over \pdr y^2}\Phi.\cr
}\eeqs
The wavefunction $\Phi$ must satisfy the condition of equivariance under the
 constant gauge transformations:
\beq
	\Phi(gWg^{-1},x,y)=g\Phi(W,x,y)g^{-1}.
\eeq
Furthermore, it should  satisfy the periodicity condition
\beq
	\Phi(W,x+2\pi,y+2\pi)=W^{-1}\Phi(W,x,y)W.
\eeq

On such functions, the above  hamiltonian is in fact translation invariant,
 although
 the separate terms are not. This provides a nontrivial consistency check of
 our formalism: there should be no dependence on the choice of the origin once
 the condition of equivariance under the constant gauge transformations are
 imposed.

We can now exploit translation invariance and change variables to
 center of mass and relative coordinates $Z=x+y$ and $z=y-x$. The result of
 a somewhat long calculation is,
\beq
	H\Phi=-2\pi R_a^2\Phi+z R_a(\eta_{aL}+\eta_{aR})\Phi+
		|z|[c_1{\cal P}_0\Phi- c_2\Phi]-{1\over m}{\pdr^2\over \pdr z^2}\Phi-{1\over
 4m}{\pdr^2\over \pdr Z^2}\Phi.
\eeq
 The constants $c_1$ and $c_2$ depend  on the particular group $G$. For
 $G=SU(2)$, $c_1=1$ and $c_2=-\half$. Here ${\cal P}_0$ denotes the projection
 to the trivial representation.

We need to rewrite the constraints in terms of the new variables, $z$ and $Z$.
The equivariance under the full gauge group is the same, but the periodicity,
is given by
\beq
         W \Phi (W,Z+4\pi,z) W^{-1}= \Phi (W,Z,z).
\eeq
This suggests  the ansatz
\beq
         \Phi (W,Z,z) =  \chi_P (W,z) e^{iPZ}.
\eeq

If we impose the periodicity constraint we get the following eqn,
\beq
         W\chi_P (W,z)W^{-1}=e^{-4iP\pi} \chi_P(W,z).
\eeq
{}From the equivariance under the constant ( global) gauge transformations,
we know that
$$\chi_P(gWg^{-1},z)=g\chi_P(W,z)g^{-1}$$
and if  we take $g=W$, we get
$$\chi_P(W,z)=W\chi_P(W,z)W^{-1}.$$
If we now use this in the  periodicity condition, we find  that  $2P$ is an
 integer.

Now one may look for solutions to the  constraint arising from the constant
 gauge transformations. In the case $G=SU(2)$, the general solution is
\beq
	\chi_P(W,z)=f_P(W,z)+Wg_P(W,z)
\eeq
where $f$ and $g$ are complex valued  central functions of $W$:
\beq
	f_P(gWg^{-1},z)=f_P(W,z)\quad g_P(gWg^{-1},z)=g_P(W,z).
\eeq

The simplest eigenfunctions of the hamiltonian  satisfying the constraint are
 those independent of $W$: $\chi_P(W,z)=f(z)$.
This corresponds to the gauge field being in its ground state. In fact, if the
 radius of the cylinder is $R$, any other state will have an energy of order
 $\alpha R$. At least when $\alpha R$ is  large, the low lying states of the
 system will consist of wavefunctions independent of $W$.  In this case, $
 \Phi(W,z,Z)=f(z)e^{iPZ}$ and the Schrodinger equation reduces to  Airy's
 equation:
\beq
        {3\over 2}|z|f(z)-{1\over m}{d^2 \over dz^2}f(z)=E'f(z) .
\eeq
The total energy is $E'+{P^2\over m}$. The physical meaning is clear: the
quarks
 have combined to form a color singlet meson, and the gluon sector is left in
 its ground state.  The energy of such a  state will approach a constant value
 as $R\to \infty$.

The   internal energy $E'$ is determined by the boundary conditions on the
 wavefunctions. The relative coordinate $z$ takes values in the range
 $[-2\pi,2\pi]$. The self--adjointness of the hamiltonian will require that
 the wavefunctions in its domain satisfy
\beq
	{\pdr\over \pdr z}\Phi(W,z,Z)+\kappa \Phi(W,z,Z)=0\;\hbox{for}\; z=-2\pi,2\pi.
\eeq
We can also require $\Phi$ and its derivative to be continuous at $z=0$.
The constant $\kappa$ is not determined by the classical theory and must be
 picked such that the limit of infinite radius of the cylinder makes sense.

The next interesting case to study would be the  bound state of $N$ quarks
 forming a baryon in the case where the structure  group is $SU(N)$. In the
 limit $N\to \infty$ this ought to tend to a soliton. It is interesting to
study
 how this soliton co-exists with the gauge excitations. We will not pursue this
 idea here.

{\bf Acknowledgements}

We thank G. Ferretti and J. Mickelsson for many discussions on this topic.
	This work was supported in part by the US Department of Energy, Grant
No. DE-FG02-91ER40685.

{\bf References}\hfill\break

\noindent\rajeev. S.G.Rajeev, Phys. Lett. {\bf B212} 203 (1988).

\noindent\hadron. S.G.Rajeev, {\it Two Dimensional Hadron Dynamics},
in preparation.

\noindent\migdal. A.Migdal, Sov. Phys. Jept. {\bf 42} 413 (1976).

\noindent\hosotani. J.Hetrick and Y.Hosotani, Phys. Lett. {\bf B230}
88 (1989).

\noindent\witten. E.Witten, Commun. Math. Phys. {\bf 141} 153 (1991).

\noindent\blauthompson. M.Blau and G.Thompson, Int. J. Mod. Phys. {\bf
A7} 3781 (1992).

\noindent\isham. C.J.Isham, {\it Modern Differential Geometry for Physicists}
 (World Scientific Publishing Co. 1989).

\noindent\percacci. R.Percacci, {\it Geometry of Nonlinear Field Theories}
(World Scientific Publishing Co. 1989).

\noindent\narasimhan. M.S.Narasimhan and T.R.Ramadas, Comm. Math. Phys.
{\bf 67} 21 (1979).

\noindent\singer. I.M.Singer, Comm. Math. Phys. {\bf 60} 7 (1978);
Phys. Scrip. {\bf 24} 817 (1981).

\noindent\babelon. O.Babelon and C.M.Viallet, Comm. Math. Phys.
{\bf 81} 515 (1981).

\noindent\mickbook. J.Mickelsson, {\it Current Algebras and Groups}
(Plenum Press, New York, 1989).

\noindent\gribov. V.N.Gribov, Nucl. Phys. {\bf B139} 1 (1978).

\noindent\mickelsson. J.Mickelsson, Phys. Lett. {\bf 242} 217 (1990).

\noindent\langmann. E.Langmann and G.Semenoff, Phys. Lett. {\bf B296}
117 (1992); Phys. Lett. {\bf B303} 303 (1993).

\noindent\ho. C.Ho and J.Hetrick, Phys. Rev. {\bf 40} 4085 (1989).

\noindent\hetrick. J.Hetrick, Amsterdam Univ. preprint,
hep-lat/9305020.

\noindent\shabanov. S.V.Shabanov, SACLAY preprint hep-th/9308002.

\noindent\ferrari.  F.Ferrari, Munich Univ. Theo. Phys.
preprint hep-th/9310024; Munich Univ. Theo. Phys. preprint
hep-th/9310055.

\noindent\saradzhev. F.M.Saradzhev, Inst. Phys. Baku, BAKU-429 (1991).

\noindent\thompsonreview. M.Blau and G.Thompson, Lectures on 2d Gauge
Theories, ICTP hep-th/9310144; G.Thompson, Lectures on Topological
Gauge Theory and Yang-Mills Theory, ICTP hep-th/9305120.

\noindent\broda. B.Broda, Phys. Lett. {\bf 244} 444 (1990).

\noindent\fine. D.S.Fine, Comm. Math. Phys. {\bf 134} 273 (1990);
Comm. Math. Phys. {\bf 140} 321 (1991).

\noindent\eliza.  L.Chandar and E.Ercolessi, Syracuse University
 preprint hep-th/9309065.

\noindent\wittenlocal. E.Witten, J. Geom. Phys. {\bf 9} 303 (1992).

\noindent\kobayashi. S.Kobayashi and  K.Nomizu,{\it Foundations of
Differential Geometry}( Interscience Publishers, Wiley, 1963).

\noindent\monopole. T.T.Wu and C.N.Yang, Phys. Rev. {\bf D12} 3845 (1975);
A.P.Balachandran, G.Marmo, B.S.Skagerstam and A.Stern,{\it Classical
Topology and Quantum States} (World Scientific Publishing Co. 1991).

\noindent\milnor. J.Milnor and J.Stascheff, {\it Characteristic Classes}
(Princeton University Press, Princeton NJ,1974).

\bye